\documentclass{elsart}
\usepackage{graphics}
\usepackage{graphicx}
\usepackage{epsfig}
\usepackage{amssymb}

\begin{document}

\begin{frontmatter}
  
\title{Studies of systematic uncertainties in the estimation of the monocular aperture of the HiRes   
  experiment}

\author[Utah]{R.U.~Abbasi},
\author[Utah]{T.~Abu-Zayyad},
\author[LANL]{J.F.~Amman},
\author[Utah]{G.~Archbold},
\author[Utah]{K.~Belov},
\author[Montana]{J.W.~Belz},
\author[Columbia]{S.Y.~Ben~Zvi},
\author[Rutgers]{D.R.~Bergman},
\author[Utah]{S.A.~Blake},
\author[Utah]{O.~Brusova},
\author[Utah]{G.W.~Burt},
\author[Utah]{Z.~Cao},
\author[Columbia]{B.C.~Connolly},
\author[Utah]{W.~Deng},
\author[Utah]{Y.~Fedorova},
\author[Columbia]{C.B.~Finley},
\author[Utah]{R.C.~Gray},
\author[Utah]{W.F.~Hanlon},
\author[LANL]{C.M.~Hoffman},
\author[Rutgers]{G.A.~Hughes},
\author[LANL]{M.H.~Holzscheiter},
\author[Utah]{P.~H\"{u}ntemeyer},
\author[Utah]{B.F~Jones},
\author[Utah]{C.C.H.~Jui},
\author[Utah]{K.~Kim},
\author[Montana]{M.A.~Kirn},
\author[Utah]{E.C.~Loh},
\author[Utah]{M.M.~Maestas},
\author[Tokyo]{N.~Manago},
\author[LANL]{L.J.~Marek},
\author[Utah]{K.~Martens},
\author[NewMexico]{J.A.J.~Matthews},
\author[Utah]{J.N.~Matthews},
\author[Utah]{S.A.~Moore},
\author[Columbia]{A.~O'Neill},
\author[LANL]{C.A.~Painter},
\author[Rutgers]{L.~Perera},
\author[Utah]{K.~Reil},
\author[Utah]{R.~Riehle},
\author[NewMexico]{M.~Roberts},
\author[Utah]{D.~Rodriguez},
\author[Tokyo]{M.~Sasaki},
\author[Rutgers]{S.R.~Schnetzer},
\author[Rutgers]{L.M.~Scott},
\author[LANL]{G.~Sinnis},
\author[Utah]{J.D.~Smith},
\author[Utah]{P.~Sokolsky},
\author[Columbia]{C.~Song},
\author[Utah]{R.W.~Springer},
\author[Utah]{B.T.~Stokes},
\author[Utah]{J.R.~Thomas},
\author[Utah]{S.B.~Thomas},
\author[Rutgers]{G.B.~Thomson},
\author[LANL]{D.~Tupa},
\author[Columbia]{S.~Westerhoff},
\author[Utah]{L.R.~Wiencke},
\author[Rutgers,LUTH]{A.~Zech\thanksref{email}},
\author[Columbia]{X.~Zhang}

\address[Utah]{University of Utah, Department of Physics and High
Energy Astrophysics Institute, Salt Lake City, Utah, USA}

\address[LANL]{Los Alamos National Laboratory, Los Alamos, NM, USA}

\address[Montana]{University of Montana, Department of Physics and
Astronomy, Missoula, Montana, USA}

\address[Columbia]{Columbia University, Department of Physics and
Nevis Laboratory, New York, New York, USA}

\address[Rutgers]{Rutgers - The State University of New Jersey,
Department of Physics and Astronomy, Piscataway, New Jersey, USA}

\address[NewMexico]{University of New Mexico, Department of Physics
and Astronomy, Albuquerque, New Mexico, USA}

\address[Tokyo]{University of Tokyo, Institute for Cosmic Ray
Research, Kashiwa, Japan}

\address[LUTH]{LUTH, Observatoire de Paris, Meudon, France}

\collaboration{The High Resolution Fly's Eye Collaboration}

\thanks[email]{To whom correspondence should be addressed.  E-mail:
\texttt{Andreas.Zech@obspm.fr}}

\date{\today}

\begin{abstract}
   We have studied several sources of systematic uncertainty in
    calculating the aperture of the High Resolution Fly's Eye
    experiment (HiRes) in monocular mode, primarily as they affect the
    HiRes-II site. The energy
    dependent aperture is determined with detailed Monte Carlo
    simulations of the air showers and the detector response.
     We have studied the effects of changes to the
  input energy spectrum and composition used in the simulation. 
   A realistic shape of the input spectrum is used in our analysis
  in order to avoid biases in the aperture estimate due to the limited 
  detector resolution. We have examined the effect of exchanging our input spectrum 
  with a simple E$^{-3}$ power law in the ``ankle'' region.
  Uncertainties in the input composition are shown to be significant
  for energies below $\sim$ 10$^{18}$ eV for data from the
  HiRes-II detector. Another source of uncertainties
  is the choice of the hadronic interaction model in the air shower
  generator. We compare the aperture estimate for two different
  models: QGSJet01 and SIBYLL 2.1.   We also
  describe the implications of employing an atmospheric database with
  hourly measurements of the aerosol component, instead of using an
  average as has been used in our previously
  published measurements of the monocular spectra. 
\end{abstract}


\end{frontmatter}

\section{Introduction}
\label{intro}

The High Resolution Fly's Eye experiment consists of two air
fluorescence detectors (``HiRes-I'' and ``HiRes-II'') located in the
desert of Utah. HiRes observes ultra-high energy cosmic rays
indirectly through extensive air showers, i.e. cascades of
secondary charged particles, which are caused by interactions of the
primary cosmic ray particles with the earth's atmosphere. In the wake
of the air shower, 
excited nitrogen molecules emit fluorescence light in the
ultraviolet, which is collected by mirrors and projected onto clusters
of photomultiplier tubes. Detailed descriptions of the detectors can
be found in~\cite{hr_det1} and~\cite{hr_det2}.

The HiRes experiment aims at measuring the arrival directions,
composition, and flux of the most energetic cosmic rays. The two
detectors allow stereoscopic observation of air showers, which yields
the best resolution in shower geometry and cosmic ray energy. An
advantage of data analysis in monocular mode, i.e. 
separate analyses of the data
from each of the two detectors, lies in the higher
statistics that can be achieved at the high energy end due to the
longer lifetime of the HiRes-I detector, which started operation two
years before HiRes-II, in 1997. Monocular analysis
also allows an
extension of the observed energy range down to energies as low as
$\sim$10$^{17}$ eV due to the larger elevation
coverage and better time resolution of the HiRes-II detector, 
and  also due to the necessity of
triggering only one detector.  The differential flux or ``energy
spectrum'' observed in monocular mode by HiRes shows a
hardening in the flux at around $10^{18.5}$ eV, known as the ``ankle'',
and a suppression of
the flux near $10^{19.8}$ eV, at the expected energy of the GZK flux 
suppression~\cite{gr,zk}.  These results have been published 
in~\cite{hr_ankle}.

The main systematic uncertainties that are introduced in the UHECR
spectrum measurement with the HiRes detectors have been reported
in~\cite{hr_prl}. They are uncertainties in the absolute phototube
calibration ($\pm$10\%), the fluorescence yield ($\pm$10\%) and the
correction for ``missing energy'' ($\pm$5\%). 
The latter refers to the energy component that is channeled
mainly into neutrinos and does not contribute to the ionization
process. Not taking into account atmospheric effects, the uncertainty
in the energy scale is $\pm$15\%, which results in a systematic
uncertainty in the flux $J$ of $\pm$27\%.  The effect on the energy
scale of a variation of the average vertical aerosol optical depth
(VAOD) by $\pm$1 RMS value, from 0.04 to 0.06 and 0.02, has also been 
described in~\cite{hr_prl}. It is not larger
than 9\% on the average. This results in a total uncertainty in
the energy scale of $\pm$17\%. The effect of the same VAOD variation
on the aperture leads to an average atmospheric uncertainty in the
flux $J$ of $\pm$15\%.  The total systematic uncertainty in the
measured flux adds up to $31 \%$ for each of the two monocular
spectrum measurements.

In this paper, we examine additional systematic uncertainties that may
affect the calculation of the HiRes aperture
in monocular mode.  Since the aperture of an air fluorescence
detector is a function of the energy of the observed cosmic rays, it
has to be modeled carefully with detailed
Monte Carlo (MC) simulations. The HiRes MC simulation programs use libraries
of air shower profiles, generated at different energies with the
air shower simulation program
CORSIKA~\cite{corsika} and the hadronic interaction 
code QGSJet~\cite{qgsjet}, for a
realistic representation of the fluctuations in the observed charged
particle profiles. A detector response MC program simulates
the light emission process along the shower and traces the photons
through the atmosphere to the telescopes of the two detectors,
taking into account all relevant atmospheric effects. The
detector optics, electronics and trigger system are modeled in great
detail using databases that record
variable detector settings, as well as density fluctuations of
aerosols in the atmosphere.  After performing extensive
comparisons between simulated events and data, which 
allow us to verify the quality of our simulations
(see~\cite{hr_hr2}), we estimate the detector acceptance using the
ratio of accepted MC events ($\nu^{MC}$) to generated
MC events ($\mu^{MC}$)  in each energy bin. 
To correctly
simulate effects stemming from the finite energy resolution of the
detectors and their limited elevation coverage, we use a continuous
energy spectrum and a bi-modal composition based on previous
measurements as inputs to our simulation programs.

The differential flux $J$ in each energy bin is calculated as:

\begin{equation}
J(E_i)= N(E_i)\cdot \frac{1}{\Delta E} \cdot \frac{1}{C_i \cdot A  \Omega \cdot t} 
\label{eq:espec}
\end{equation}

where $N(E_i)$ is the number of observed events in the energy bin and
$\Delta E$ is the bin-width. The geometrical aperture (area $\times$
solid angle) used in generating MC events is
noted by $A \Omega$, and $t$ is the detector live-time.
Through our use of a continuous and realistic input energy spectrum,
the finite energy resolution of the detectors is taken
into account in the acceptance $C_i$ =
$\frac{\nu_i^{MC}}{\mu_i^{MC}}$.  This will be explained in the next chapter.
In the following, we refer to the product of the constant $A \Omega$
and the acceptance as (instantaneous) aperture.

We will first consider the effects of varying the input energy
spectrum on the 
calculated aperture and thus on the measured spectrum, in
Section~\ref{einput}. In
Section~\ref{hadmod}, we examine the implications of
exchanging the hadronic interaction model in the air shower generator.
For this study, we replace the QGSJet model, which is used in our standard
spectrum measurement, with the SIBYLL
model~\cite{sibyll}. The effect of a variation of the assumed input
composition on the measured spectrum is presented in Section~\ref{compinput}.
Another systematic uncertainty that can affect
the aperture estimate of the experiment is that due 
to variations in the aerosol component of the atmosphere. For
the analysis of the monocular spectra, we used an average atmospheric
description based on measurements with laser systems that are
installed at each detector site~\cite{hr_atmos1}.  In 
Section~\ref{atmos}, we re-analyze the HiRes-II monocular data
with a database containing hourly measurements of the aerosol
component of the atmosphere and compare it to the average description
in our standard analysis.  
Although the systematic studies presented in
this paper have been carried out with simulation and reconstruction
tools of the HiRes-II analysis, their results are 
applicable to the HiRes-I spectrum measurement as well.

\section{Input Energy Spectrum Bias}
\label{einput}

The calculation of the cosmic ray energy spectrum from the measured
energy distribution of events is a problem of unfolding the true
spectrum of cosmic rays at their arrival at the earth's
atmosphere from the distortions introduced by
the detector. The energy distribution provided by the
detector is a convolution of the true spectrum with the detector
response, i.e. the efficiency of the detector and its finite
resolution.  Following the discussion in
G. Cowan's \emph{Statistical Data Analysis}~\cite{cowan}, the
problem of unfolding can be stated in the following way:

\begin{equation}
\nu_i = \sum_{j=1}^M R_{ij} \mu_j
\label{eq:responsematrix}
\end{equation}  

Here, the true energy spectrum and the measured spectrum are divided
into $M$ energy bins; $\mu_j$ is the number of events in bin $j$ of
the true histogram, $\nu_i$ the expectation value of the number of
events in bin $i$ of the measured histogram. $R_{ij}$ is the response
matrix, which describes the detector response in each energy bin.
Off-diagonal elements in $R_{ij}$ are due to the
limited resolution of the detector, which distributes a fraction of
events from a certain energy bin over adjacent bins.

The most straightforward way of determining the real event
distribution $\mu_j$ from the measured values 
is to calculate the response matrix and apply its
inverse to the measured distribution.
Determining the response matrix requires knowledge of the
detector resolution and acceptance, as well as a good
estimate of the true spectrum.  However, as Cowan shows, even with a
complete knowledge of $R_{ij}$, this method is not applicable in
most cases since it leads to
large variances in the unfolded histogram, when
the resolution is large compared to the
bin-width. These variances arise due to the Poisson distribution of
the observed data around the expectation values $\nu_i$.

In practice, the ``method of correction factors'' can be applied for
the deconvolution of the measured spectrum. This is the method
used in our analysis. The estimator $\hat{\mu}_i$ for the true
spectrum is written as:

\begin{equation}
\hat{\mu}_i = C_i^{-1} \cdot n_i
\label{eq:cfmethod}
\end{equation}

where $n_i$ are the observed data and $C_i^{-1}$ are multiplicative
correction factors for each energy bin. These correction factors are
determined with MC simulations of both the physical
model under study and the complete measurement process. They are just
the inverse of the acceptance estimate $C_i$, which is given by the ratio 
of accepted over generated events in the MC in each energy bin:

\begin{equation}
C_i = \frac{\nu_i^{MC}}{\mu_i^{MC}} = \frac{\nu^{MC}(E_i)}{\mu^{MC}(E_i^\prime)}
\label{eq:correctionfactor}
\end{equation}

The distribution of accepted events $\nu^{MC}$ is evaluated at the reconstructed
energies $E_i$, whereas the distribution of generated events $\mu^{MC}$ is given as
a function of the true (input) energies $E_i^\prime$.
Calculation of the expectation value for the estimator $\hat{\mu}_i$
provides an expression for the bias of the method of correction
factors.

\begin{eqnarray}
\displaystyle E[\hat{\mu}_i] = C_i^{-1} \cdot E[n_i] = C_i^{-1} \cdot \nu_i = \mu_i + \left( C_i^{-1} - \frac{\mu_i}{\nu_i} \right)\nu_i 
\label{eq:bias1}
\end{eqnarray}

The bias in the estimator $E[\hat{\mu}_i]$ is given by the
last term of Equation~\ref{eq:bias1}.  It goes to zero as the estimated acceptance,
$C_i$, approaches the true acceptance
of the experiment, $\frac{\nu_i}{\mu_i}$.  The more realistic
the assumptions that go into the MC simulation are,
the smaller the bias will be. One can estimate the bias by varying the
model used in the simulation.

We have calculated an estimate of the bias by varying the assumed true
energy spectrum that is used as an input to the MC.  It is useful to
rewrite the term that describes the bias in the following way:

\begin{equation}
b_i = \left( C_i^{-1} - \frac{\mu_i}{\nu_i}
\right)\nu_i 
= \left( \frac{\nu_i}{\mu_i} \cdot C_i^{-1} - 1
\right)\mu_i = (R-1)\mu_i
\label{eq:bias2}
\end{equation}

The bias as a fraction of the real spectrum $\mu_i$ can thus be
calculated from the ratio $R$ of the true to the
estimated acceptance.  For our bias estimate, we assumed the true
acceptance $\frac{\nu_i}{\mu_i}$ to be the result of a simulation
using our best estimate of the input energy spectrum. The estimated
acceptance $C_i$ was 
calculated using a simple $E^{-3}$ power law for the input energy spectrum.

\begin{figure}[thb]
\begin{center}
\includegraphics[height=30.0pc]{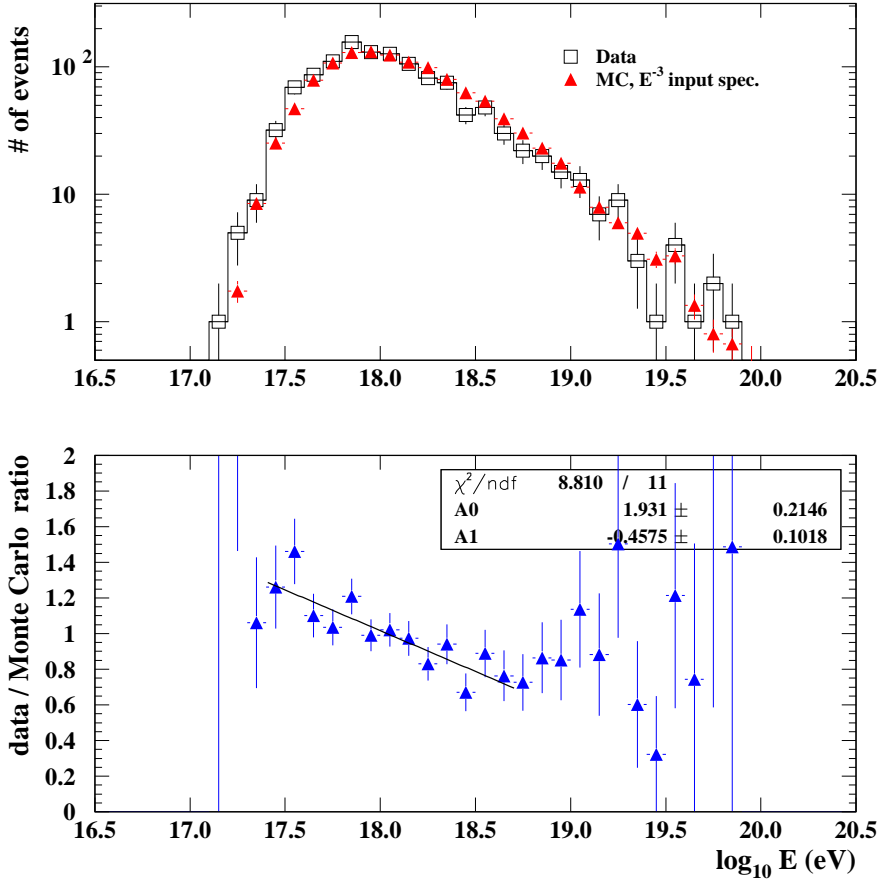}
\end{center}
\vspace{-0.5pc}
\caption[Data-MC Comparison: energy distribution for $E^{-3}$ input
spectrum in MC.]  {
  The top panel shows the energy distribution of HiRes-II data.
    The histogram and square points show the actual data; the
    triangles show the MC distribution, \emph{assuming an $E^{-3}$
      input spectrum,} normalized to the same total number of
      events.  The bottom
    panel shows the ratio of the data to MC distributions 
    from the top plot with a linear fit below the ``ankle''.
  }
\label{fig:inpspec1}   
\vspace{0.5pc}
\end{figure}

Figure~\ref{fig:inpspec1} shows the measured energy distribution
for data and a MC simulation assuming an $E^{-3}$
input spectrum. About one
third of the HiRes-II data used in our
monocular spectrum measurement published in~\cite{hr_ankle} have
been included in this comparison. As can be seen from
the distributions, and more clearly from the ratio plot (lower
panel, data divided by MC), this choice of the input spectrum
is not very good.  The ratio is not
flat because the assumed input spectrum does not have a break
(``ankle''). Thus, if one normalizes data and MC to the same total number of 
events, the fraction of MC events is too small at low energies and too large at
higher energies. We have used this MC set to calculate the biased acceptance
estimate $C_i$.

\begin{figure}[thb]
\begin{center}
\includegraphics[height=30.0pc]{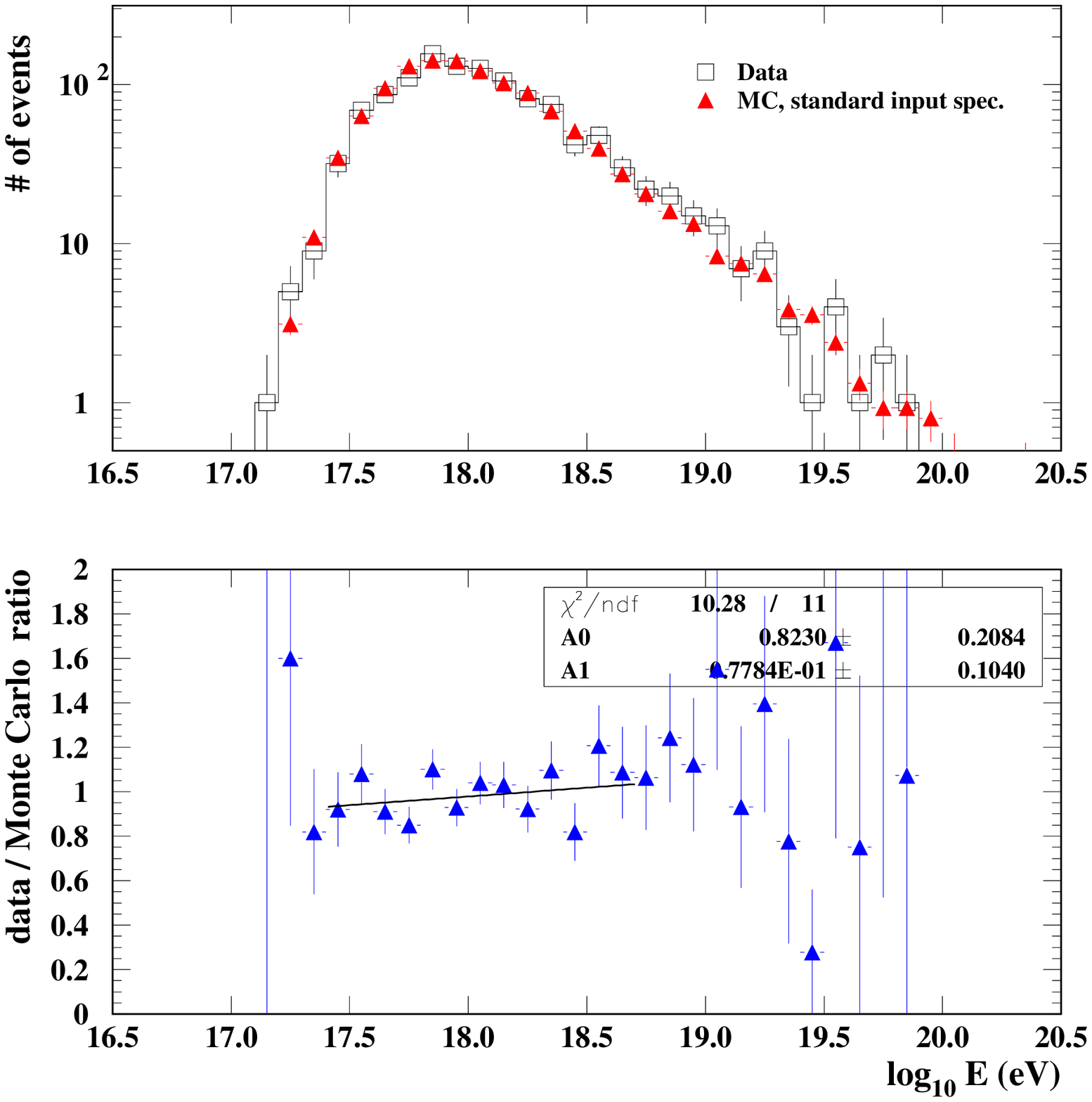}
\end{center}
\vspace{-0.5pc}
\caption[Data-MC Comparison: energy distribution for {\it Fly's Eye} input
spectrum in MC.] {
  The top panel shows the energy distribution of HiRes-II data.
    The histogram and square points show the actual data; the
    triangles show the MC distribution, \emph{assuming an input
      spectrum based on broken power law fits to the Fly's Eye stereo
      spectrum~\cite{flyseye} and HiRes-I spectrum~\cite{hr_prl}},
      normalized to the same total number of events.
    The bottom panel shows the ratio of the data to MC distributions
    from the top plot with a linear fit below the ``ankle''. }
\label{fig:inpspec2}   
\vspace{0.5pc}
\end{figure}

The bias has been corrected in Figure~\ref{fig:inpspec2}: instead of the
$E^{-3}$ spectrum, we now use a fit to the Fly's Eye stereo
spectrum~\cite{flyseye} to determine the shape of the input spectrum
below the ``ankle'', and a linear fit to the HiRes-I spectrum for higher energies.
The spectral index of this input spectrum is -3.01 between 10$^{16.5}$ eV and
10$^{17.6}$ eV, -3.27 between 10$^{17.6}$ eV and 10$^{18.7}$ eV, and -2.80 above
10$^{18.7}$ eV. In this study, the position of the ``ankle'' is assumed to be at 
10$^{18.7}$ eV, corresponding to the first results of the HiRes spectrum measurements 
published 
in~\cite{hr_hr2}. Our more recent result with higher statistics in the HiRes-II data-set 
observes the ``ankle'' at 10$^{18.5}$ eV. The linear fit
above the ``ankle'' is extended to the highest energies. The good agreement
between data and MC shows that this choice of input spectrum is closer to the true spectrum
$\mu_i$ given that the MC simulates all other aspects of the experiment well, which was
shown in~\cite{hr_hr2}.  This MC set is used to estimate the true
acceptance $\frac{\nu_i}{\mu_i}$.

\begin{figure}[thb]
\begin{center}
\includegraphics[height=30.0pc]{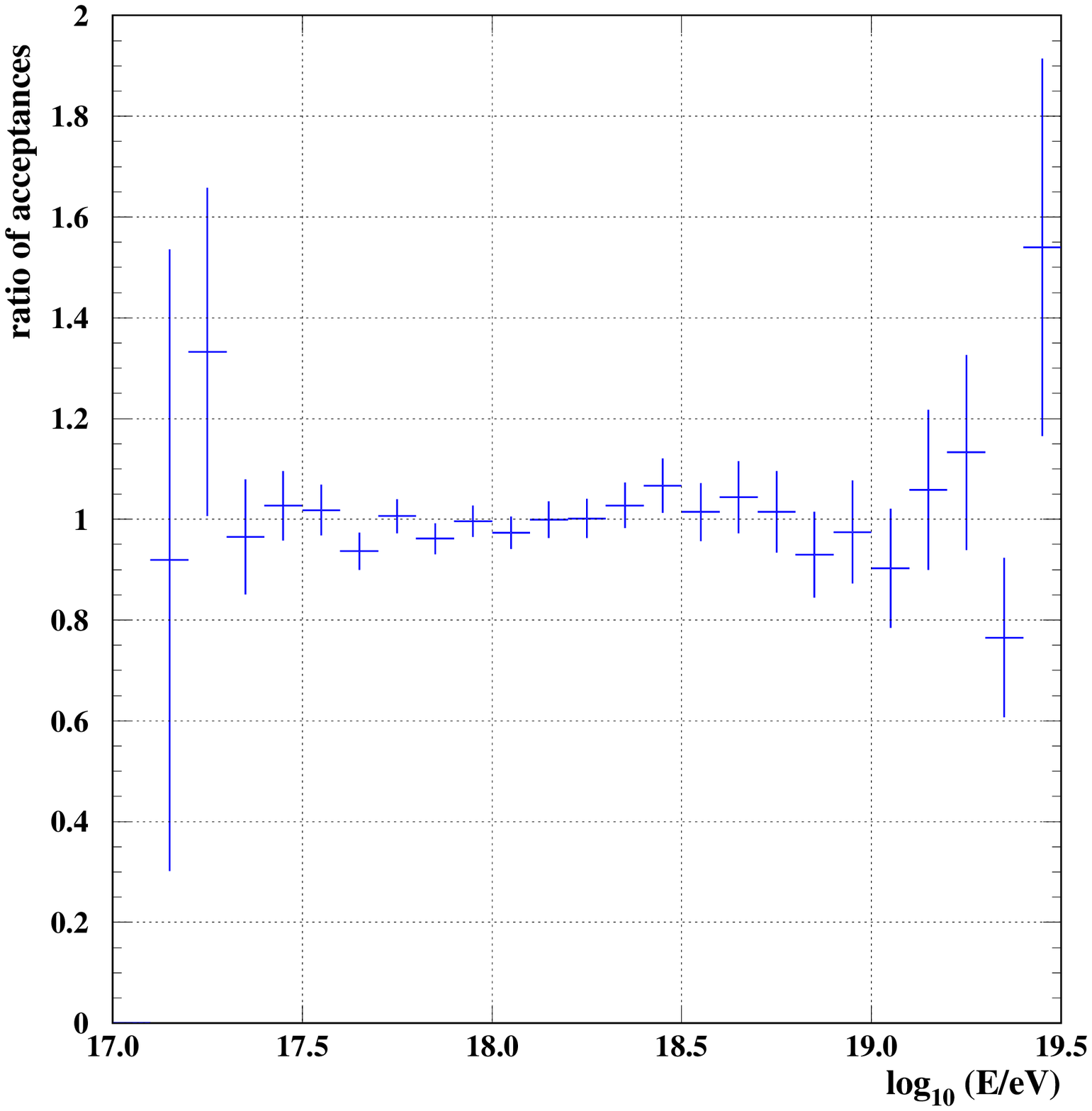}
\end{center}
\vspace{-0.5pc}
\caption[Ratio of acceptances for different input spectra.]
{Ratio of the acceptance calculated using
    an input spectrum based on the Fly's Eye and HiRes-I
    measurements, as described in the text, to the acceptance calculated
    assuming an $E^{-3}$ input spectrum.}
\label{fig:bias}   
\vspace{0.5pc}
\end{figure}

The nearly flat ratio of the data and MC distributions in
Figure~\ref{fig:inpspec2} means that $\frac{\nu_i}{\nu_i^{MC}}$ is
approximately constant if one chooses a realistic input
spectrum.  In this case, according to Equation~\ref{eq:bias1}, the energy
dependence of the expectation value for the true spectrum
$E[\hat{\mu}_i]$ is approximately given by the input spectrum
$\mu_i^{MC}$. Any differences in the unfolded spectrum can be fed back
into the MC and will improve the agreement between the energy
distributions in real and simulated data, thus reducing the bias in
the spectrum calculation with the updated MC simulation.

The bias we avoid by including the ``ankle'' feature in the input
spectrum can be derived from Figure~\ref{fig:bias}
which shows the ratio $R$ of acceptances for the two MC
simulations. A kink is visible in the ratio plot in the ``ankle''
region, even though the effect is very small. The ratio increases from
$\sim$0.97 at 10$^{18}$ eV to $\sim$1.07 at 10$^{18.5}$ eV and then decreases
to $\sim$0.94 at 10$^{18.8}$ eV. This bias is due to the 
limited energy resolution of the detectors which spreads event energies
over neighboring bins. It should be noted that the same random number seeds
have been used in both MC sets to reduce statistical fluctuations in the ratio
plot. In our standard analysis, we smooth the calculated acceptance by replacing 
the acceptance histogram with an appropriate fitting function. The remaining
statistical fluctuations are then taken into account in the spectrum measurement.
For this study, we have not applied any smoothing procedures.

In our analysis of the HiRes-II spectrum, we used initially an E$^{-3}$ input 
spectrum, which
was soon replaced by a spectrum with the shape of the Fly's Eye measurement~\cite{flyseye}.
This input spectrum was used for our publication of the monocular spectra~\cite{hr_prl},~\cite{hr_hr2}. 
Our measurement of the position of the ``ankle'' and the spectral index
above the ``ankle'', as described above, was used for our updated spectrum publication~\cite{hr_ankle}.
Adjusting the input spectrum to feature the ``ankle'' shape has helped us to significantly
improve the agreement in several comparison plots between data and MC events, which we use
to evaluate our simulation programs. The effect on the acceptance is rather small, as seen 
in Figure~\ref{fig:bias}. Replacing the E$^{-3}$ input spectrum with an ``ankle'' shape 
led to a variation in the acceptance of less than 10\%. Any 
further adjustments of the exact shape of the ``ankle'' had negligible effects on 
the spectrum.

Thus far, we have not included the observed flux suppression~\cite{hr_ankle} 
above 10$^{19.8}$ eV in our input energy spectrum.

\section{Hadronic Interaction Model Uncertainty}
\label{hadmod}

When simulating events, we
read profiles of charged particles from a library of air showers and
simulate the light generation, propagation, and detector response for
different shower geometries. This ``shower library'' contains a large
collection of profiles in steps of 5 g/cm$^2$ vertical atmospheric
depth at several fixed energies and at a zenith
angle of 45$^{\circ}$.  The shower profiles were
generated with
CORSIKA for proton and iron primaries and fitted with 
the Gaisser-Hillas function~\cite{ghfunc}:

\begin{equation}
N(X)=N_{max} \left( \frac{X-X_0}{X_{max}-X_0} \right)^{(X_{max}-X_0)/
  \lambda} \exp((X_{max}-X)/ \lambda)
\label{eq:ghfit}
\end{equation}

The three fit parameters, X$_{max}$ (the atmospheric slant depth
at shower maximum), N$_{max}$ (the number of charged
 particles at shower
maximum) and $\lambda$, were written into the library files for
each shower to characterize its profile. $X_0$ was fixed at -60
g/cm$^2$.  Correlations between the mean 
values of the fit
parameters and the logarithm of the shower energy are used to scale
shower profiles from the fixed energies provided in the library to the
continuous energy spectrum required in the detector response MC, as
described in~\cite{hr_hr2}.  Our analysis uses Gaisser-Hillas fits to
the charged particle profiles of air showers to estimate the
ionization energy of observed and simulated showers. The integral over
the fitted profile is multiplied by a mean ionization loss rate,
derived from simulations with CORSIKA to be 2.19
MeV/(g cm$^2$)~\cite{song}. Before the
integration is carried out on MC events, the particle profile
has to be adjusted for a fraction of 10\% of the primary energy that
is lost due to cuts on particles with energies below preset thresholds
in CORSIKA.  We then also have to determine the ``missing energy'',
which does not contribute to the ionization
 process, by comparing the
estimated ionization energy of the library showers to their known
total energy.  A correction for the ``missing energy'' is added to the
reconstructed energies of all simulated and real events.

Newer CORSIKA versions provide directly information on the energy
deposit profile of the air shower, but in this study we want to apply the
same algorithms used in our published analysis of the monocular spectra.
We have verified that our method yields results consistent with the energy
deposit profiles.

\begin{figure}[thb] 
 \begin{center}
 \includegraphics[height=30.0 pc]{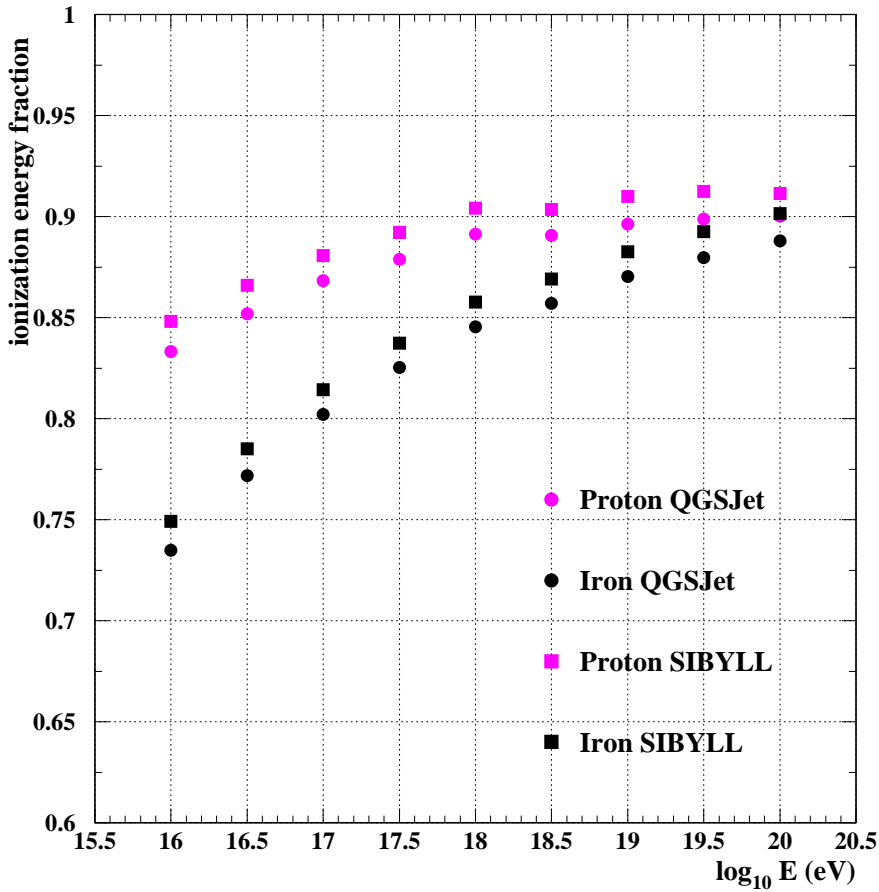}
  \end{center}
\vspace{-0.5pc}
  \caption[Ionization energy estimates for QGSJet and SIBYLL.]
  {Estimates of ionization energy fraction derived from shower
    profiles. The ratio of ionization energy to total
    energy is shown for proton (upper points, in magenta) and iron
    primaries (lower points, in black) versus the logarithm of the
    total energy.  The squares are results from simulations with
    SIBYLL, the circles correspond to QGSJet.}
  \label{fig:eion}
 \vspace{0.5pc} 
\end{figure}

The physics contributing to the electromagnetic
component of the air shower is well understood and described in detail
by the EGS code~\cite{egs} within the CORSIKA program framework. The
main uncertainty in the air shower simulation stems from our limited
knowledge of the initial hadronic interactions, which take place at
energies by far exceeding those that can be observed in the
laboratory. In order to get an estimate of the influence of those
uncertainties on the calculated aperture, we have generated two
``shower libraries'' using two different hadronic interaction models.
CORSIKA 5.61 with QGSJet01~\cite{qgsjet} has been used in our standard analysis.  
For this study, we generated an updated shower library with
CORSIKA 6.022 and QGSJet01. The second model we 
chose was SIBYLL 2.1~\cite{sibyll} (with the same CORSIKA version).  
Differences between the two models can be
found in their predictions of the particle multiplicity, inelasticity
and the extrapolations of the hadron-air cross-section to ultra-high
energies.  A detailed comparison is given in~\cite{hadmod}. With
regard to the charged particle profiles we are interested in,
differences can be seen in the mean X$_{max}$ values and the
elongation rates ($\frac{d<X_{max}>}{d \log{(E)}}$). SIBYLL showers,
especially in the case of proton primaries, have on average
larger X$_{max}$ values and a slightly different elongation rate, as
can be seen in Figure~\ref{fig:composition} (in the next chapter).  Another difference between
the two models is shown in Figure~\ref{fig:eion}.
Our estimates of the ionization energy fraction are roughly 2\% larger when 
using SIBYLL, as compared with QGSJet.

For the estimation of the detector aperture, we follow the same
procedure with each of the two hadronic interaction models: 
the fit parameters for the ``shower libraries'' are taken from
Gaisser-Hillas fits to the shower profiles. We found that the
Gaisser-Hillas function with three parameters describes accurately the
particle profiles for showers generated with either of the two
hadronic interaction models. 
We determine the MC input composition from HiRes/MIA and HiRes stereo
 measurements of the mean X$_{max}$ as a function of the cosmic ray energy.
 At a given energy, the mean X$_{max}$ of the air shower distribution is correlated
 with the average mass of the primary cosmic ray flux. 
We assume a simple bi-modal composition of protons and iron nuclei and determine
a proton fraction by comparing the data points with the iron and proton estimates given
by the two models.
This procedure will be described in more detail in the following chapter.

\begin{figure}[thb] 
 \begin{center}
\includegraphics[height=30.0 pc]{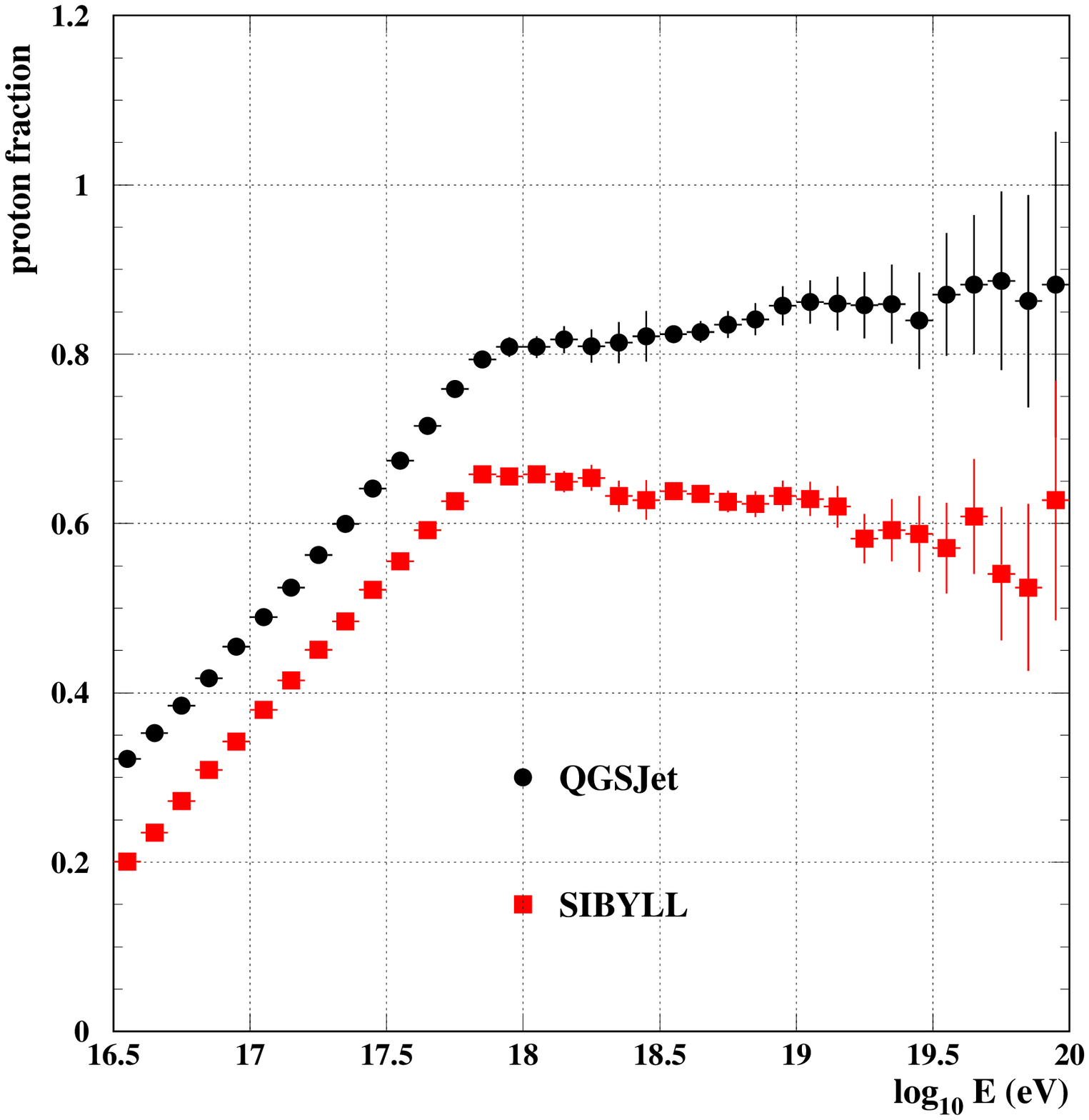}
  \end{center}
\vspace{-0.5pc}
  \caption[Proton fraction of the MC input composition for QGSJet and SIBYLL.]
  {Proton fractions used for the MC input composition
    for QGSJet and SIBYLL.}
  \label{fig:pfrac}
 \vspace{0.5pc} 
\end{figure}

\begin{figure}[thb] 
\begin{center}
\includegraphics[height=30.0pc]{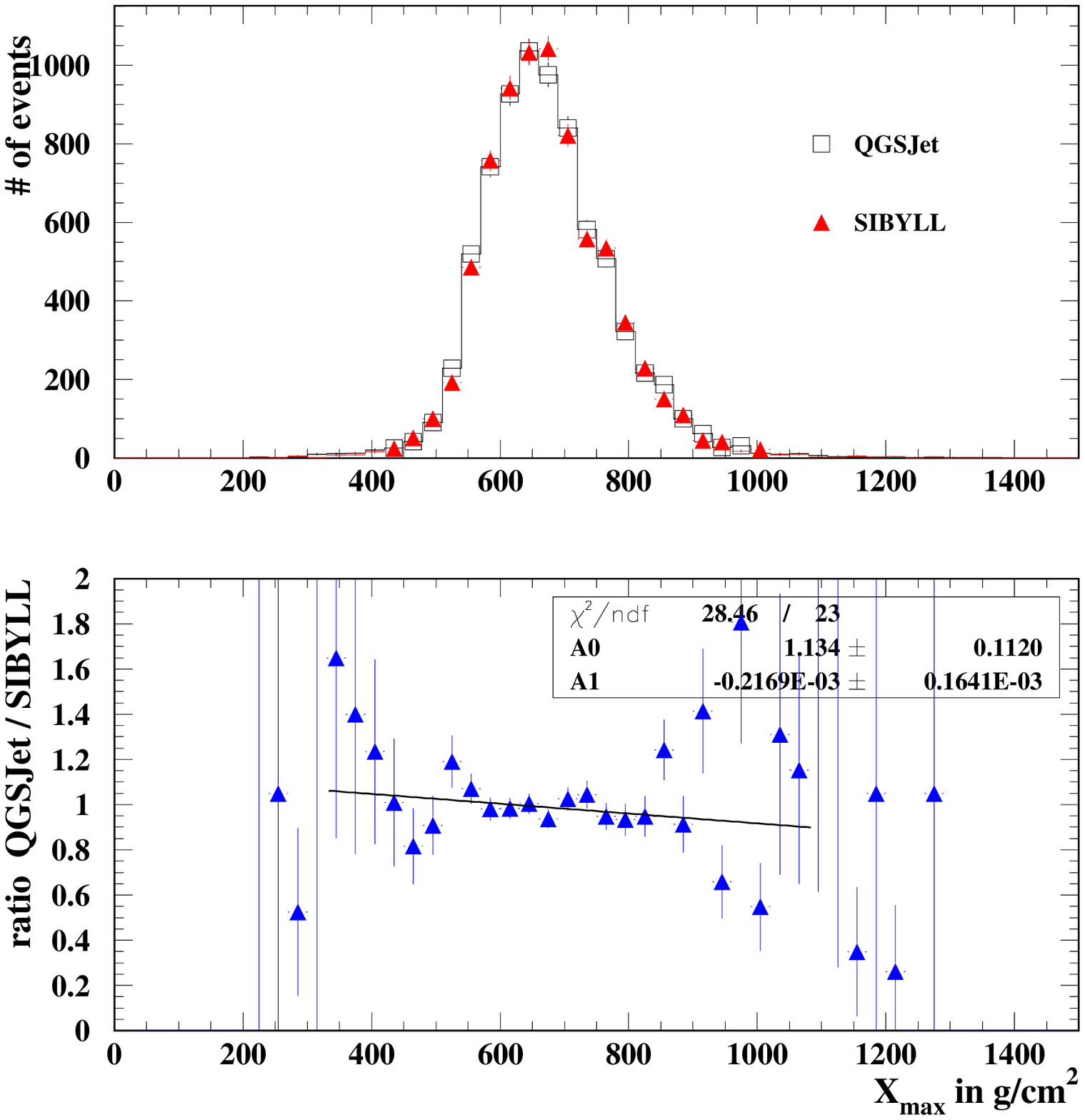}
\end{center}
\vspace{-0.5pc}
\caption[QGSJet-SIBYLL Comparison: Distributions of reconstructed
X$_{max}$ .]  { 
  The top panel shows the MC distributions of reconstructed
    X$_{max}$ using QGSJet (squares) and SIBYLL (triangles).  
    The two distributions have been normalized to cover the same
    area.  The bottom panel shows the ratio of the two distributions
    shown in the top panel.}
\label{fig:xmax_q_s}   
\vspace{0.5pc}
\end{figure}

Since the mean X$_{max}$ values for
pure proton showers are larger in the case of SIBYLL, we had to
re-calculate the proton fraction that corresponds to the data points
and adjust the input composition to contain a larger fraction of iron
showers. The difference in the proton fractions used as
input to the MC for the two models are shown
in Figure~\ref{fig:pfrac}.  The X$_{max}$ distributions of reconstructed
MC events that passed all our quality requirements are shown in
Figure~\ref{fig:xmax_q_s} for the two models.  The close agreement of the
distributions for the QGSJet and SIBYLL simulations 
demonstrates that we place simulated showers at the
same distribution of atmospheric depths for 
either model.

In both cases, we determine the ``missing energy'' from a comparison
of the total shower energy to the integral of the
shower profile that has been multiplied by the mean
ionization loss rate. Instead of applying an average correction for
proton and iron showers, we determine the correction for the fraction
of simulated proton and iron showers that were accepted in our detector 
response simulation and successfully reconstructed.

\begin{figure}[thb] 
 \begin{center}
 \includegraphics[height=30.0 pc]{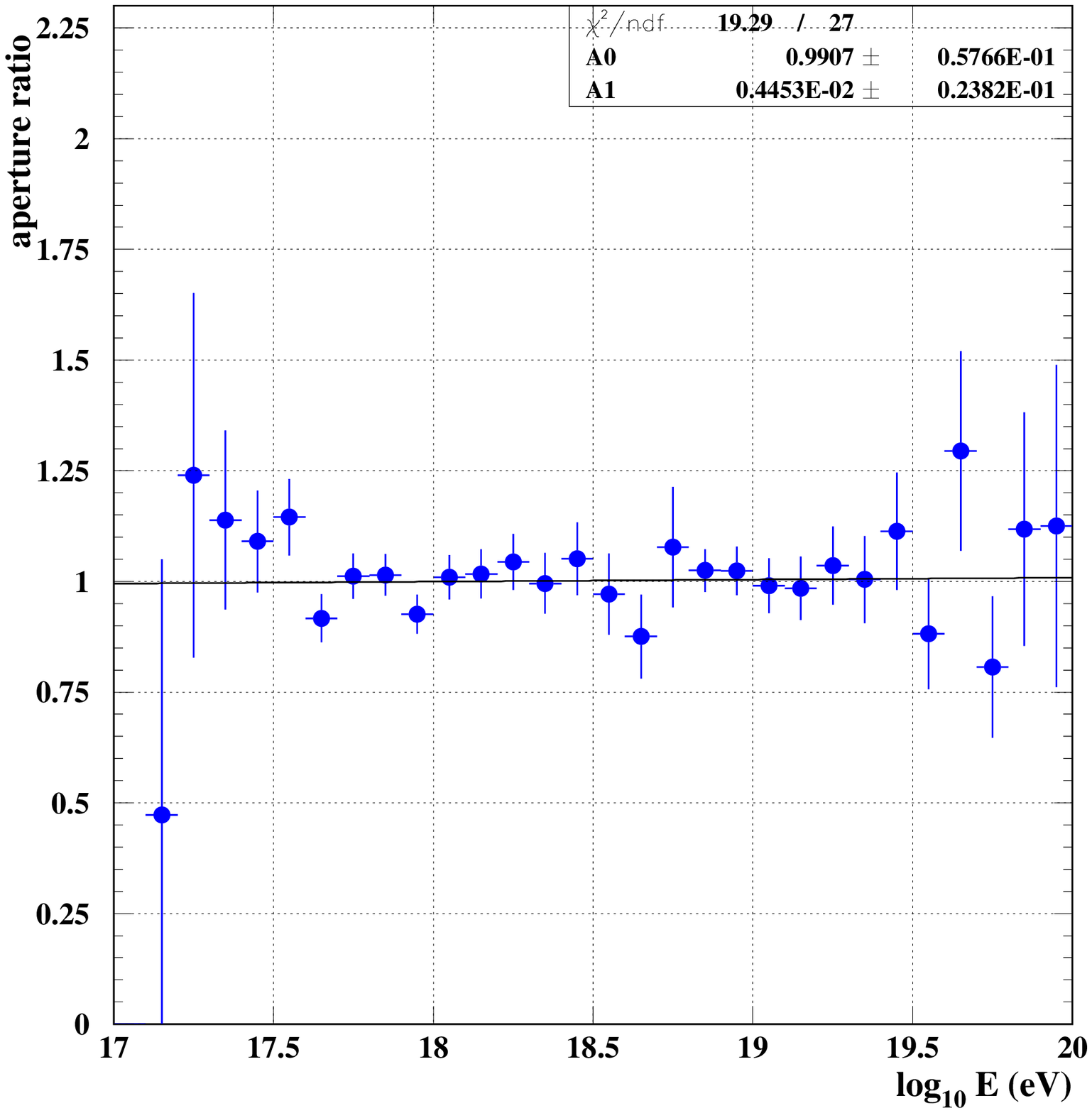}
  \end{center}
\vspace{-0.5pc}
  \caption[Ratio of apertures for SIBYLL and QGSJet.]
  {Ratio of apertures using SIBYLL 2.1 (numerator) and
    QGSJet01 (denominator) with linear fit.}
  \label{fig:hadratio}
 \vspace{0.5pc} 
\end{figure}

Using the same analysis procedure for each of the two hadronic
interaction models, we did not find any significant differences in our 
extensive set of comparisons between distributions of data and
simulated events with the two MC sets.  Figure \ref{fig:hadratio} shows
the ratio of the apertures that result from simulations using the
QGSJet and SIBYLL libraries of air showers. 
No smoothing algorithms have been applied to the calculated acceptances.
The same random number seeds were used for the two MC sets to reduce statistical
fluctuations.
Both the normalization, which is consistent with 1, and the zero slope of 
the fit to this ratio show that the effect is negligible compared to the 
statistical uncertainties in our data-set. We thus find
that if we apply our procedure to estimate the detector aperture in a
consistent way, the result does not depend on the chosen
hadronic interaction model. This is important since the models are
continuously evolving. The latest version of the QGSJet model (QGSJet02) 
for example has been shown to generate mean X$_{max}$ values closer 
to the predictions of SIBYLL 2.1~\cite{ostapchenko}.

\section{Input Composition Uncertainty}
\label{compinput}

The fraction of air showers initiated by light and heavy (i.e.\ 
proton and iron) cosmic rays used in our MC simulation is
determined from composition measurements by the
HiRes/MIA~\cite{hrmia1}~\cite{hrmia2} and HiRes
Stereo~\cite{hr_stecomp} experiments. Air fluorescence detectors like
HiRes can measure the cosmic ray composition as a function
of energy in a statistical way. The atmospheric depth,
X$_{max}$, at which an extensive air shower
reaches its maximum size depends not only on the energy but also on
the mass of the primary cosmic ray particle. On the average, heavy nuclei interact
higher in the atmosphere than light nuclei of the same energy. Nuclei
break up into fragments, each of which generates a sub-shower, thus
distributing the initial energy over several cascades. (In a highly simplified
picture, an iron shower is approximated by the superposition of 56
proton showers of a factor of 1/56 smaller energies.)  Unfortunately, statistical
fluctuations between shower profiles are large and do not allow an
event-by-event determination of the cosmic ray
composition.  Only the X$_{max}$ distribution for
a given energy bin can be measured and compared to
model predictions of protons and iron nuclei.

\begin{figure}[thb] 
 \begin{center}
\includegraphics[height=30.0 pc]{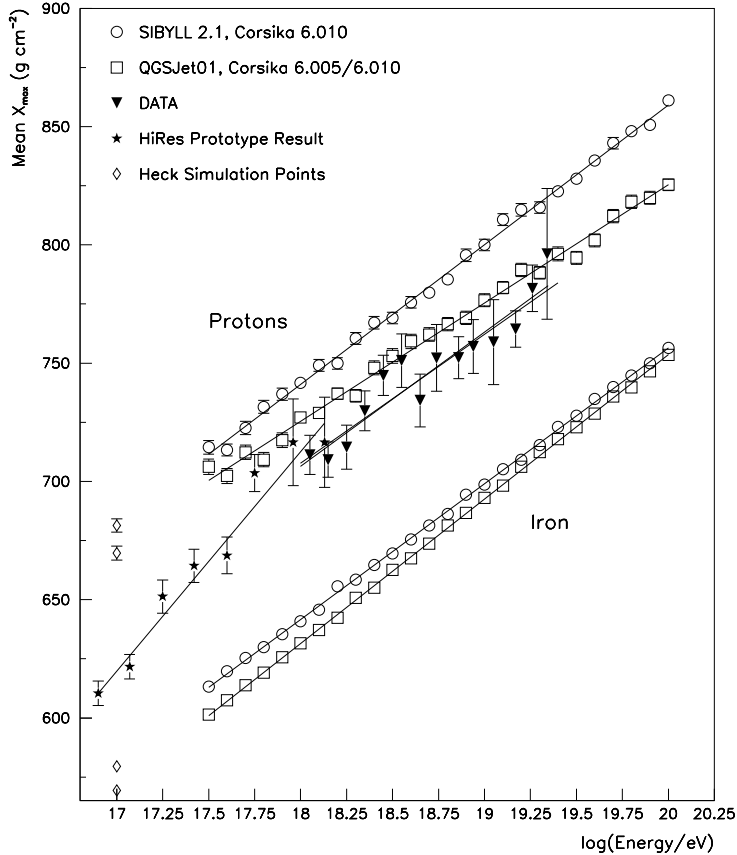}
  \end{center}
\vspace{-0.5pc}
  \caption[HiRes/MIA and HiRes Stereo Composition.]
  {X$_{max}$ vs. the logarithm of the energy from HiRes/MIA (stars)
    and HiRes Stereo (triangles) measurements. The predictions for
    QGSJet and SIBYLL are shown for comparison. The diamonds are
    simulation points calculated by D. Heck. The best fit to
    each data set is shown. A fit to
    the 76\% of HiRes Stereo data with hourly
    atmospheric corrections is included as well. This fit has a
    slightly steeper slope. The figure is taken
    from~\cite{hr_stecomp}.}
  \label{fig:composition}
 \vspace{0.5pc} 
\end{figure}

Figure~\ref{fig:composition} shows the measured mean X$_{max}$ together
with the pure proton and iron estimates from different models.
We have re-interpreted the HiRes/MIA measurement by comparing it against 
the QGSJet01 model~\cite{qgsjet}. We use linear fits to the HiRes/MIA and HiRes 
Stereo points to determine an energy dependent proton fraction $f(E)$ by comparing
the fits with the simulated iron and proton lines of QGSJet01. In this simple bi-modal
model, we derive $f(E)$ from the distance of the fitted data points to the proton and
iron lines. A data point on the proton line would have an $f$ of 1, whereas 
a data point in the middle between the two simulation lines would have an $f$ of 0.5. 
The derived proton fraction is 0.45 at 10$^{17}$ eV, 0.80 at 10$^{17.85}$ eV and 1.0
at 10$^{20}$ eV.
The proton fraction we determine from the measurements and insert into our
MC simulations depends thus on a specific interaction model. 
However, by using the same model as a reference for the input
composition and for the simulation of air showers, we generate events
with the measured X$_{max}$ distribution independently of
the chosen model, as was shown in the previous section.

\begin{figure}[thb] 
 \begin{center}
\includegraphics[height=30.0 pc]{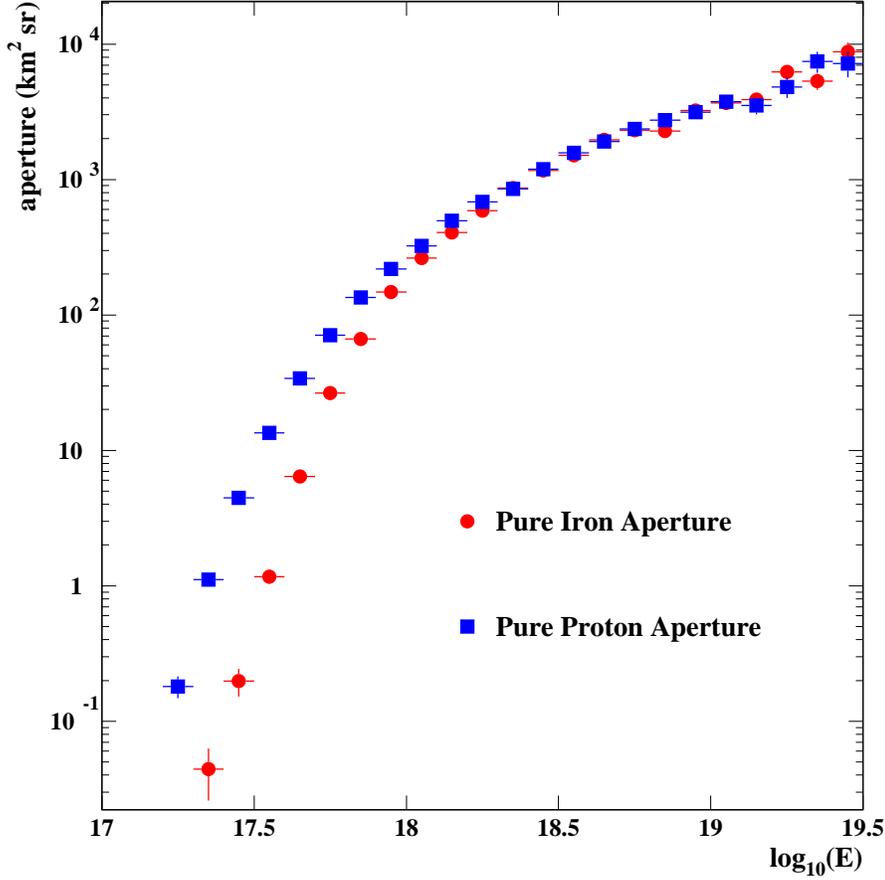}
  \end{center}
\vspace{-0.5pc}
  \caption[Apertures for pure iron and pure proton MC sets.]
{Apertures for pure iron and pure proton MC sets.}
  \label{fig:exp_prot_iron}
 \vspace{0.5pc} 
\end{figure}

Here, we investigate the effect of a
change in the measured mean X$_{max}$ on our
aperture estimate. For this study, we have used the same MC programs
as in our standard analysis, i.e. the HiRes-II detector response simulation 
and CORSIKA 5.61 with QGSJet01 for the air shower generation.
The difference in the estimated aperture
between a MC set with only iron events and a set with only proton
events can be seen in Figure~\ref{fig:exp_prot_iron}. At the low energy
end of the spectrum, the aperture for iron cosmic rays is lower
because iron showers develop higher up in the atmosphere and are more
likely to lie above the HiRes-II elevation coverage (3$^{\circ}$ to
31$^{\circ}$) than proton showers.  This leads to larger differences
between the two apertures at lower energies.  For energies above $\sim
10^{18}$ eV, where showers are on average farther away from
the detector, no significant difference is seen in the
aperture for iron and proton showers.

Using the pure iron and
pure proton apertures, we have calculated the effect of a change in
the assumed proton fraction on the aperture estimate.
The proton fraction $f(E)$ is defined as the
ratio of generated proton showers over the sum of generated proton and
iron showers in the MC: 
$f(E) = \mu_p(E) / (\mu_p(E) + \mu_{fe}(E))$.  The
acceptance for a MC set with a proton fraction $f$ in a given energy
bin is:

\begin{equation}
a_f = \frac{\nu_p + \nu_{fe}}{\mu_p + \mu_{fe}} = \frac{\nu_p(1+\nu_{fe}/\nu_p)}{\mu_p/f}=
a_p (R+f \cdot (1-R))
\label{eq:amix}
\end{equation}

Here $\nu_p$ and $\nu_{fe}$ are the accepted, i.e.\ triggered and well
reconstructed, proton and iron events, respectively;
$a_f$ and $a_p$ are the acceptances for a MC set with proton fraction
$f$ and 1, respectively.  $R$ is given by the ratio of the acceptances
for pure iron and pure proton MC sets $\left( \frac{ \nu_{fe} / \mu_{fe} }
 { \nu_p / \mu_p } \right)$. This ratio can be
determined directly from the two curves shown in
Figure~\ref{fig:exp_prot_iron}, since the apertures are just the acceptances
multiplied by a constant factor, the geometrical aperture $A \Omega$. 
With $R$ known, Equation~\ref{eq:amix} yields
the acceptance $a_f$ for a given proton fraction $f$ in 
a given energy bin.
 
It can be seen from Figures~\ref{fig:exp_prot_iron} and~\ref{fig:composition} that 
systematic uncertainties in the aperture due to uncertainties in the proton fraction are 
only of concern at the low energies covered by the HiRes/MIA measurement.We have calculated
the systematic uncertainty in the proton fraction $f(E)$ from the relevant uncertainties 
in energy and X$_{max}$ quoted in the HiRes/MIA PRL paper~\cite{hrmia2}.

Sources for energy uncertainties in HiRes/MIA are the detector calibration ($< $5\% 
uncertainty in energy) and the aerosol component of the atmosphere ($< $10\% 
uncertainty in energy). A 10\% uncertainty in the fluorescence yield is common to both 
HiRes and HiRes/MIA, and is therefore not included in our calculation. Since both
experiments use the same assumptions on the fluorescence yield, a potential error 
in this parameter would induce the same bias in the reconstructed energies of HiRes/MIA
and HiRes. It would thus not change the shape of the aperture. Given the measured 
elongation rate of 93 g/cm$^2$~\cite{hrmia2}, the uncertainties in energy from calibration
and atmosphere, added in quadrature, contribute $<$4.4 g/cm$^2$ to the uncertainty in 
X$_{max}$. 

The quoted uncertainty in X$_{max}$ of roughly 25 g/cm$^2$ due to the
calculation of the Cherenkov fraction is also common to the two experiments and is thus
not relevant for our calculation. 
Since the same assumptions on the Cherenkov light beam are made in the
HiRes and HiRes/MIA analysis, a potential bias in the HiRes/MIA reconstruction would be corrected 
in the HiRes detector simulation before the calculation of the acceptance.
In other words, the HiRes MC simulation positions showers on the average at the same height
where they were seen by HiRes/MIA.
A recent study of the fluctuations of the molecular density profile using radio sonde 
data shows a small discrepancy with the standard model used in both HiRes/MIA 
and HiRes~\cite{fedorova}. This introduces an additional uncertainty in X$_{max}$ 
of $<$10 g/cm$^2$.

Since the separation between the proton and iron lines in the QGSJet01 model is 
$\sim$100 g/cm$^2$, the uncertainties in X$_{max}$ of 4.4 g/cm$^2$ from the energy 
measurement and of 10 g/cm$^2$ from the molecular density fluctuations translate
to $\sim$4.4 \% and $\sim$10 \% uncertainty in the proton fraction $f(E)$, respectively. 
Finally, one has to add a $\sim$3\% uncertainty coming from the linear fit to the HiRes/MIA 
data that is used to parameterize the proton fraction in the simulation programs.
Those uncertainties in the HiRes/MIA measurement of the mean X$_{max}$ that
translate into uncertainties in the acceptance add then up to $\sim$11\% of the
proton fraction.

\begin{figure}[h] 
 \begin{center}
\includegraphics[height=30.0 pc]{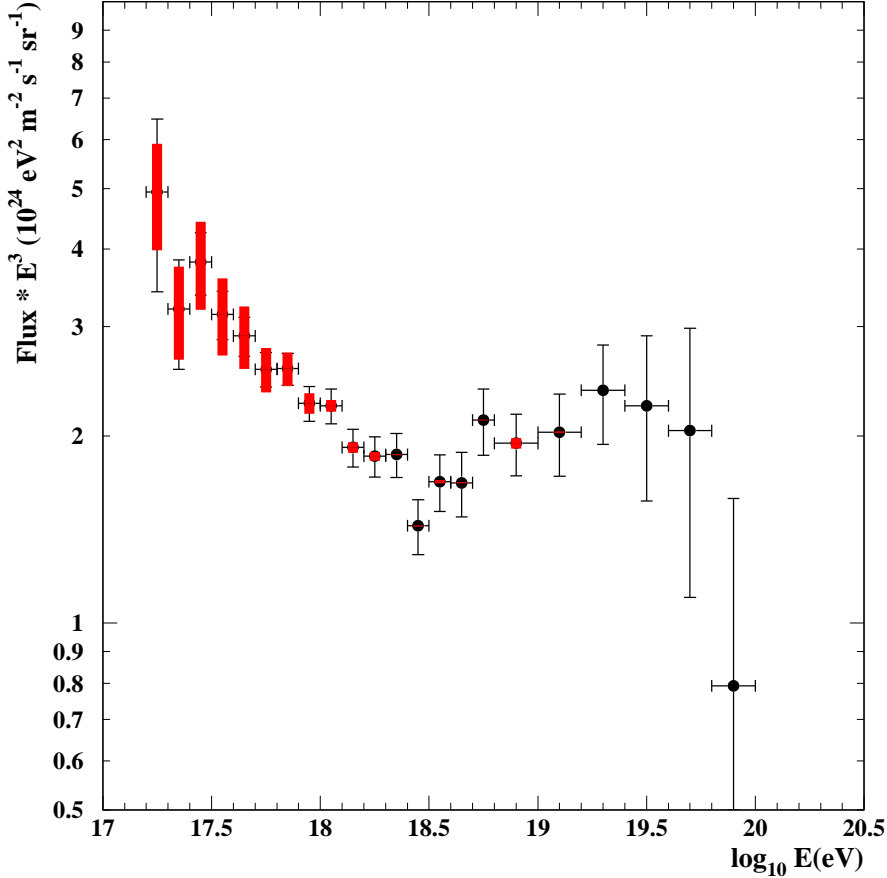}
  \end{center}
\vspace{-0.5pc}
  \caption[Systematic uncertainties due to input composition.]
  { HiRes-II energy spectrum with systematic
    uncertainties (thick error bars) corresponding to a $\pm$11\%
    change in the proton fraction of the MC. Data included in the
    spectrum were recorded between December 1999 and September 2001.}
  \label{fig:compsys}
 \vspace{0.5pc} 
\end{figure}

With the help of equation~\ref{eq:amix}, we have calculated the variation in the
aperture $a_f(E)$ for a variation in $f(E)$ of $\pm$11\%. A change in the aperture 
translates directly into a change of the measured spectrum.
The uncertainties in the spectrum from a $\pm$11\%
variation in the proton fraction are shown in Figure~\ref{fig:compsys} as
thick error bars.  At the low energy end of the HiRes-II spectrum, the
systematic uncertainties from the input composition are comparable to
the statistical uncertainties in the spectrum. In the absence of a more 
precise composition measurement in the HiRes/MIA energy range, it will thus be
difficult to observe the feature of the ``second knee'' 
\cite{2knee_fe},\cite{2knee_hrmia},\cite{2knee_akeno},\cite{2knee_yakutsk}, even with 
better statistics in the HiRes-II data. Above $\sim$10$^{18}$
eV, the effect on the aperture estimate becomes negligible.

\section{Atmospheric Uncertainties}
\label{atmos}

In the ultraviolet (UV) fluorescence range observed by the HiRes
telescopes (310-400 nm), light attenuation comes from mainly two sources:
Rayleigh scattering on air molecules, and
absorption and scattering on aerosols.  While Rayleigh scattering is well
  understood and does not vary much over time, attenuation by aerosols
  can change over short time ranges and has to be monitored during the
  data taking process. We use a system of steerable lasers, one
at each of the two detector sites, to measure light
attenuation by aerosols on an hourly basis. The
UV laser at each site fires shots in a regular pattern of varying
geometries, which are observed from the other
site. The vertical aerosol optical depth (VAOD) can be measured from the
detected light of vertical shots. Horizontal attenuation length (HAL)
and scattering phase function due to aerosols can be measured from 
the light scattered into the telescopes under different angles from horizontal 
shots. The wavelength of the laser is 355 nm, close to the 357 nm fluorescence 
line. We account for the wavelength dependence of aerosol scattering in our 
simulation and reconstruction programs.

For the monocular spectra published in~\cite{hr_ankle} and~\cite{hr_prl}, we
used a measurement of the average VAOD and HAL in our analysis. This
was necessary since the steerable laser system became fully
operational only two years after the HiRes-I detector had started
taking data. For consistency in the analysis of the two monocular
spectra, we have thus used a single average measurement, while
applying strict cuts on the selection of clear nights that were included in
the spectrum. Here, we repeat the analysis for HiRes-II using a
database with hourly entries of the measured VAOD and HAL instead 
of the average values.

\begin{figure}[thb] 
  \begin{center}
  \includegraphics[height=30.0pc]{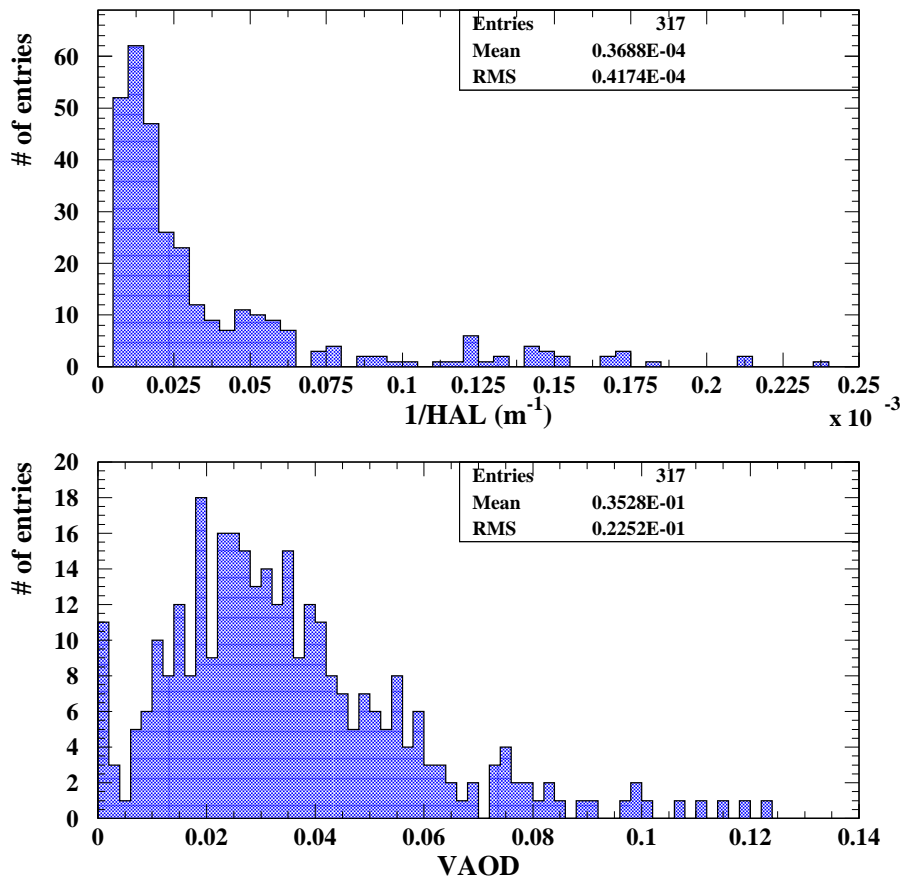}  
  \end{center}
  \vspace{-0.5pc}
  \caption[VAOD and inverse HAL for all three datasets.]
  {Hourly VAOD and inverse HAL measurements for the selection of clear
    nights in the HiRes-II analysis (December 1999 to September 2001).}
 \label{fig:atmo_ds123}
\vspace{0.5pc} 
\end{figure}

The inverse HAL and VAOD distributions for the selected clear nights
that went into this analysis are shown in Figure~\ref{fig:atmo_ds123}.  In
the atmospheric database generated for the HiRes-II analysis, entries
are available for 82 out of the 122 nights that were selected for the
spectrum calculation.  When measurements of both 
parameters were not available, a seasonal average
was assumed.  The averages found in this analysis represent a slightly
clearer atmosphere than the values used in the standard monocular
analyses ($<$HAL$>$ = 25 km and $<$VAOD$>$ = 0.04 were used for our
published spectra). This difference is due to the data normalization
method applied here, which was not used in the original analysis
(see~\cite{hr_atmos1} and~\cite{hr_atmos2}). The averages determined 
here ($<$HAL$>$ = 27 km and $<$VAOD$>$ = 0.035) are nevertheless well 
within the quoted uncertainties of the averages used in the monocular analyses.

\begin{figure}[thb] 
 \begin{center}
 \includegraphics[height=25.0 pc] {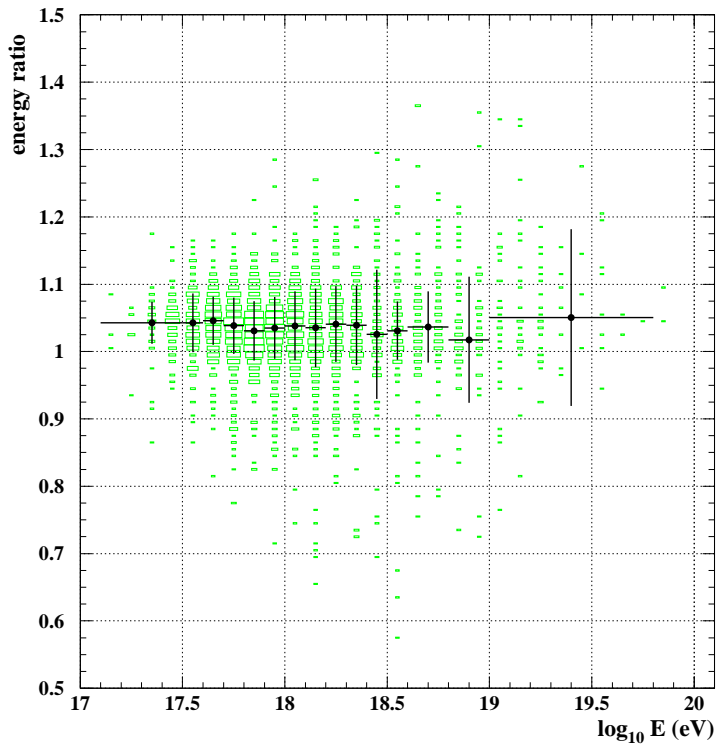} 
  \end{center}
\vspace{-0.5pc}
  \caption[Difference in reconstructed energies due to atmospheric variations.]
  {Difference in reconstructed energies due to atmospheric variations.
    The energy of each data event has been reconstructed using both    
    the atmospheric database and 
    an average atmosphere.  Shown is the ratio of the two
    reconstructed energies (average over database) versus the
    energy reconstructed using
    the atmospheric database. The sizes of the boxes are proportional to 
    the number of entries. The
    markers represent the mean and width of a Gaussian fit to the
    distribution in each energy bin.}
  \label{fig:ediff_vs_e}
  \vspace{0.5pc}
\end{figure}

In order to study the effect of variations in the aerosol component of
the atmosphere on the reconstructed energies, we have analyzed the
HiRes-II data from December 1999 to September 2001 using the
atmospheric database. Since all events were reconstructed
both with the atmospheric database and with the
average atmospheric values, the ratio of the energy estimates can be
calculated for each event. The distribution of those
ratios is shown in Figure~\ref{fig:ediff_vs_e} as a function of the
energy reconstructed using the
atmospheric database. A Gaussian fit has been applied to the
distribution in each energy bin. The points represent the Gaussian
means, the error bars the standard deviations. The
energies reconstructed with database are on the average 
4\% smaller.  This is due to the slightly clearer
atmosphere determined with the improved analysis method for the VAOD
values that went into the database. There is no significant energy 
dependence in the ratio of the energies.

\begin{figure}[thb] 
 \begin{center}
 \includegraphics[height=30.0 pc]{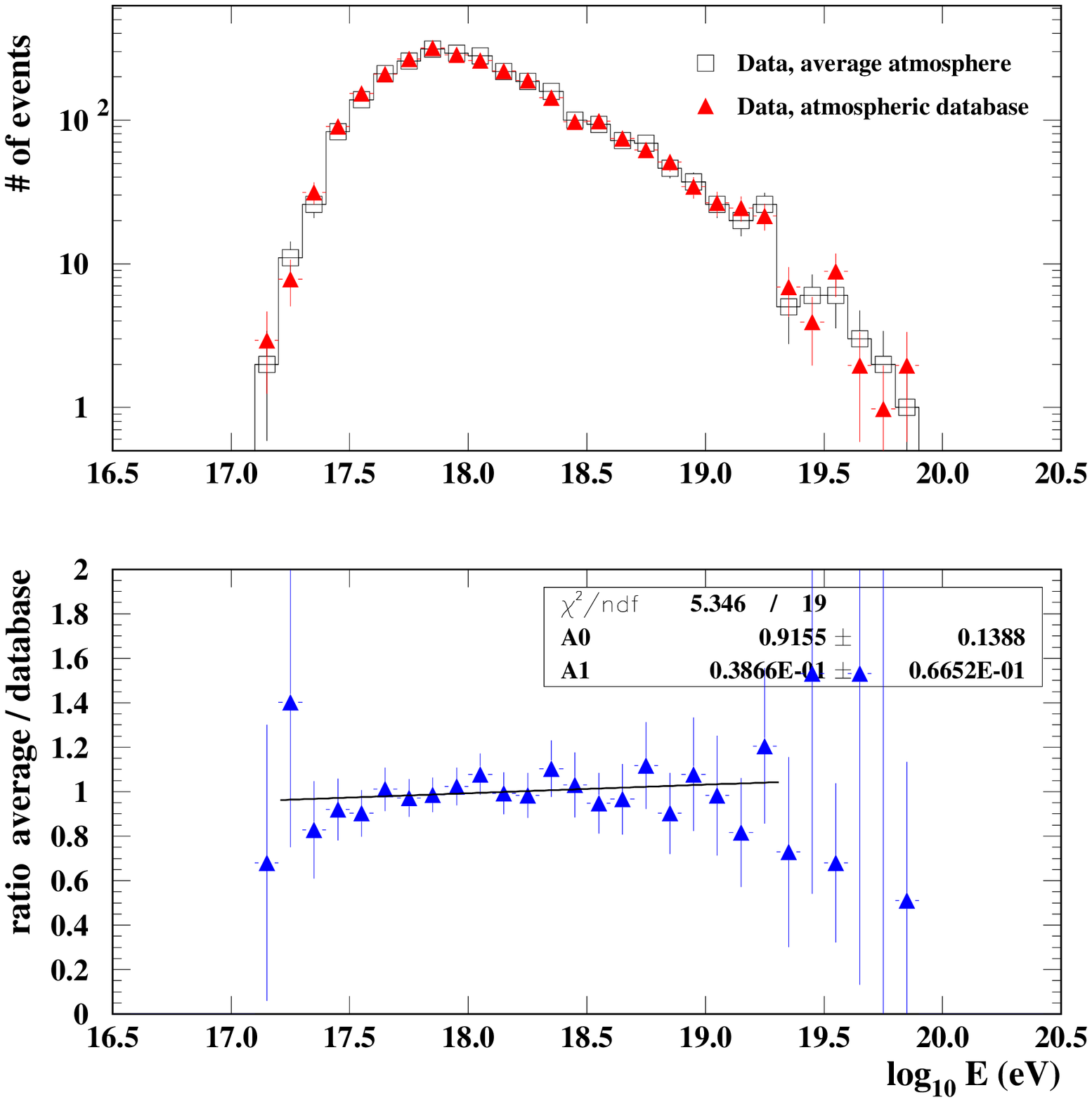} 
  \end{center}
\vspace{-0.5pc}
  \caption[Energy distribution with atmospheric database.] { 
  The top panel shows the energy distribution of HiRes-II data
    analyzed using the atmospheric database (triangles) and analyzed using average
    atmospheric parameters (squares). The bottom panel shows the
    ratio of the two distributions with the average atmosphere case as
    the numerator and the atmospheric database case as the
    denominator.}
  \label{fig:ds123_ave_adb}
 \vspace{0.5pc}
\end{figure}

\begin{figure}[thb] 
 \begin{center}
  \includegraphics[height=30.0 pc]{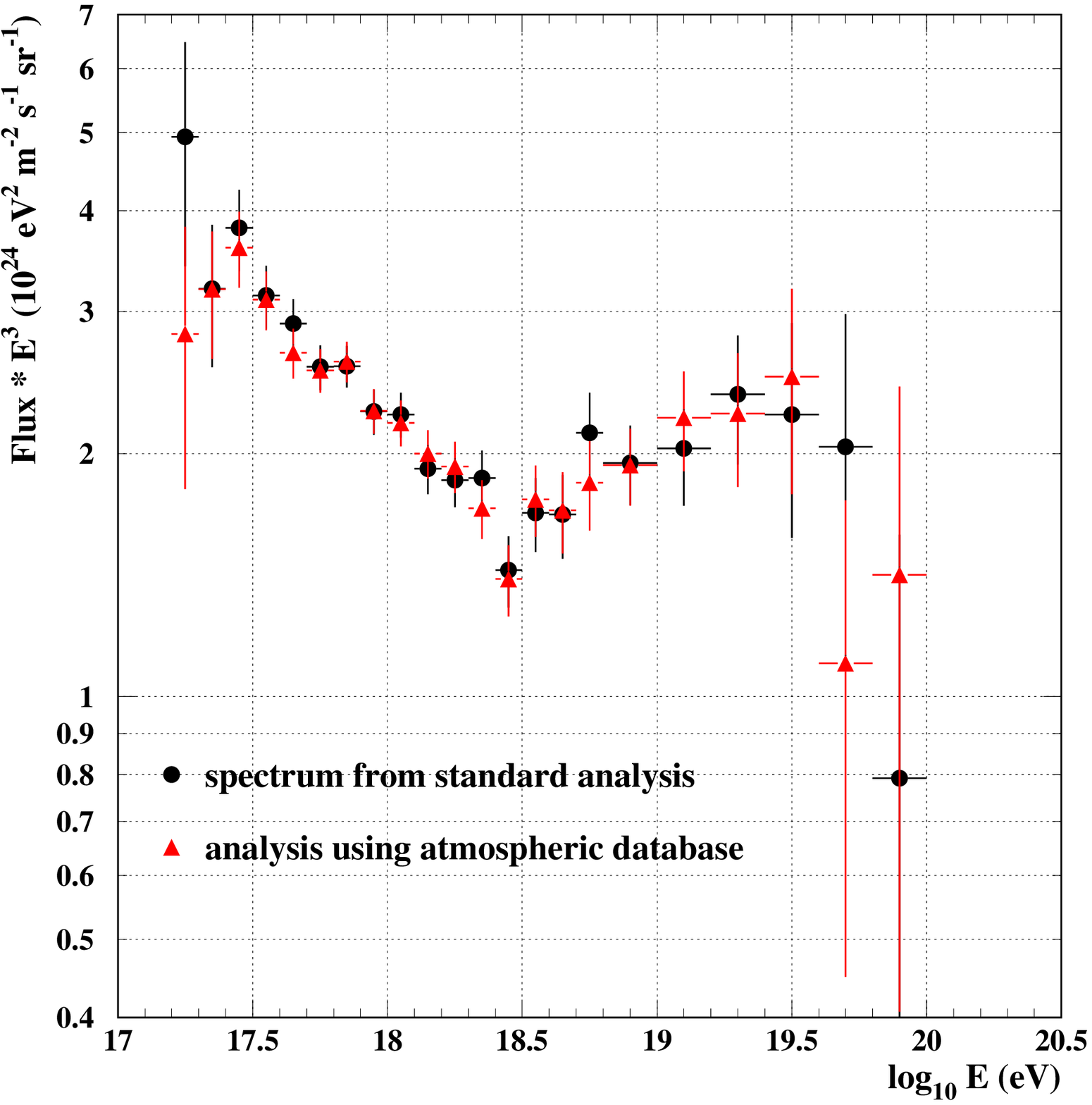}
  \end{center}
\vspace{-0.5pc}
  \caption[Standard spectrum and spectrum with atmospheric database.]
   {The HiRes-II spectrum from our standard analysis is shown in circles. 
    The spectrum resulting from an analysis that was using the atmospheric database in
   the determination of the aperture and in the reconstruction of the data is shown in triangles.}
  \label{fig:spec_std_adb}
 \vspace{0.5pc}
\end{figure}

We then examined the effect of the small shift in reconstructed energies
on the distribution of events over energy bins $N(E_i)$, which goes into the 
spectrum calculation.The histograms for the two energy reconstructions, using the
average atmosphere and the atmospheric database, can be seen in 
Figure~\ref{fig:ds123_ave_adb}. The 4\% shift in energy is too small to cause
a significant effect in the event distribution given our bin-size, which is 
adapted to the data statistics.It should be noted that the difference between
the two distributions is here even aggravated by the fact that two slightly different 
versions of our reconstruction software were used. The histogram for
the average atmosphere is the exactly same as in the calculation of the published 
HiRes-II spectrum, which permits a direct comparison of this figure with 
Figure~\ref{fig:spec_std_adb}, wheres for the reconstruction with database we had to use a
slightly updated version of our analysis software. (For Figure~\ref{fig:ediff_vs_e}, only 
the updated version was used.) 

\begin{figure}[thb] 
 \begin{center}
  \includegraphics[height=30.0 pc]{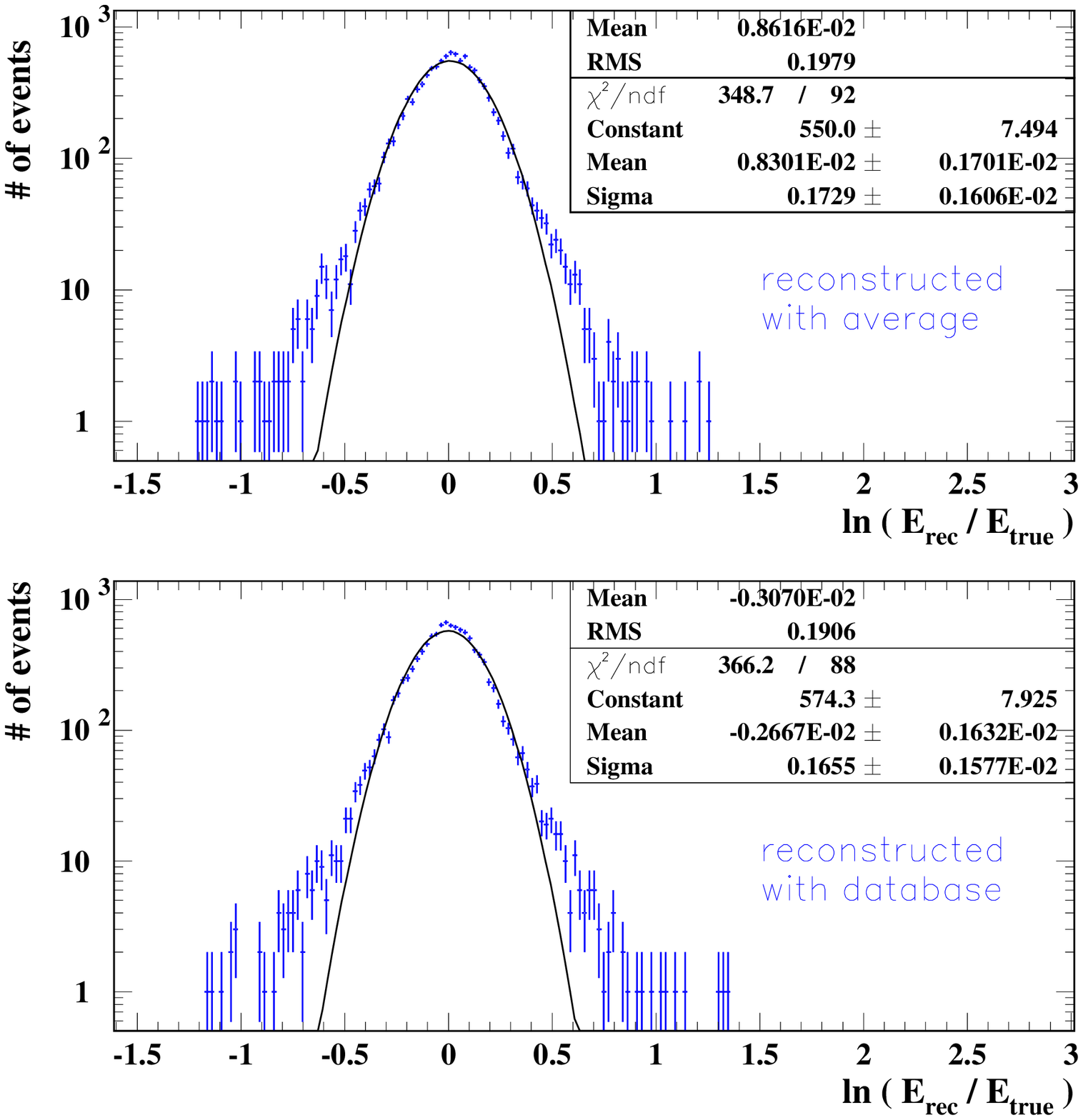}
  \end{center}
 \vspace{-0.5pc}
  \caption[Energy resolution with atmospheric database.]
   {The energy resolution is shown as the natural 
   logarithm of the ratio of reconstructed
    over true (input) energy.  The top panel shows the
      resolution in a MC set generated using the atmospheric database and
      reconstructed with seasonally averaged atmospheric parameters. 
     For the ratio of reconstructed over true energy (not 
      its logarithm) one obtains a width of 21.9\% (RMS) and a $\sigma$ of 
      the Gaussian fit of 18.9\%. 
       The bottom panel shows the resolution in
      the same MC set when analyzed using the database. 
      The RMS width and $\sigma$ of the ratio are 21.0\% and 18.0\%, 
           respectively.
      }
   \vspace{0.5pc}    
  \label{fig:res_ave_adb}
 \vspace{0.5pc}
\end{figure}

The effect of atmospheric variations on the energy resolution can be
studied with simulated events. We have generated a MC set with about
four times data statistics using the atmospheric database. The MC
events have been reconstructed in two different ways: first with the
seasonally averaged atmospheric values and then 
with the database for nights when
atmospheric data were available. A comparison of the resolution
estimates is shown in Figure~\ref{fig:res_ave_adb}.
It should be noted that the two plots use a logarithmic scale, 
hence the tails in the distributions are very small. There is no significant
difference between the tails of the two distributions. Only the width differs
by a small amount. Reconstructing the MC
events with seasonally averaged atmospheric values instead of using the
atmospheric database widens the resolution by 0.9\% (both in RMS and $\sigma$
of the Gaussian fit).

Finally, we have analyzed the effect of using the atmospheric database
rather than the measured average on the energy spectrum.
We have calculated the acceptance from a MC set that was
generated and reconstructed with the database.
As in our standard analysis, we have applied a smoothing procedure to
minimize statistical fluctuations in the acceptance. The remaining statistical
uncertainties of the smoothed acceptance are taken into account in the
measurement of the spectrum. The HiRes-II data were
also reconstructed with use of the atmospheric database. In this way,
the hourly measurements of atmospheric variations were included in
every step of the analysis. The energy spectrum 
resulting from this analysis is compared to the published spectrum, which uses
the nominal averages of VAOD (0.04) and HAL (25 km), in
Figure~\ref{fig:spec_std_adb}. The result for $J E^3$ does not vary by 
more than $\pm$15\% at any energy, except for the first and the last two bins, 
where statistics in the data are limited.
The difference between the two spectra in the last two bins is due to a
single event that has shifted up in energy to the last bin when reconstructed
using the atmospheric database, as can be seen from Figure~\ref{fig:ds123_ave_adb}. 

The focus of the study presented in this chapter was on the difference between reconstruction
results with an atmospheric database and with an average atmosphere.We
have also examined the systematic uncertainty in the determination of the atmospheric
parameters that describe the aerosol distribution (HAL and VAOD). We have compared
the atmospheric database used in this study with an independent result from an 
analysis of laser shots with different geometries, which were reconstructed with 
independently developed software.
The reconstructed energies one obtains with this independent database are on the 
average 5\% lower than the values presented here. The average difference remains
smaller than 7\% at the highest energies.

\section{Conclusions}
\label{conclusions}

None of the sources of possible systematic uncertainties we have
studied here contribute significantly to our published estimate
of the systematic uncertainty. The bias introduced by using an E$^{-3}$
power law instead of a more realistic spectral shape is not very significant
in the ``ankle'' region. Our
calculated aperture is sensitive to the assumed input composition for
energies below $\sim$10$^{18}$ eV for HiRes-II.  By using a measured
composition as an input to our simulation programs, our analysis does
not depend on the assumed hadronic interaction model.  For
the 17 month period tested here, the description of the aerosol
density using an hourly database does not cause any significant
differences in the spectrum, when compared with an average atmosphere.
We also found no significant changes in
the reconstructed energies for the time period under study.

\section*{Acknowledgements}

This work is supported by US NSF grants PHY-9321949, 
PHY-9322298, PHY-9904048, PHY-9974537, PHY-0098826, 
PHY-0140688, PHY-0245428, PHY-0305516, PHY-030098,  
and by the DOE grant FG03-92ER40732. We gratefully 
acknowledge the contributions from the technical 
staffs of our home institutions. The cooperation of 
Colonels E.~Fischer and G.~Harter, the US Army, and 
the Dugway Proving Ground staff is greatly appreciated.


\begin{thebibliography}{00}

\bibitem{hr_det1} T. Abu-Zayyad {\it et al.}, Proceedings of the 26th
  International Cosmic Ray Conference {\bf 5}, 349 (1999)

\bibitem{hr_det2} J. Boyer, B. Knapp, E. Mannel and M. Seman, Nucl.
  Inst. and Meth. {\bf A 482}, 457 (2002)
  
\bibitem{gr} K.~Greisen, Phys. Rev. Lett. {\bf 16}, 748 (1966).
  
\bibitem{zk} G.T.~Zatsepin and V.A.~K'uzmin, Pis'ma Zh. Eksp. Teor.
  Fiz.  {\bf 4}, 114 (166) [JETP Lett. {\bf 4}, 78 (1966)].

\bibitem{hr_ankle} R. U. Abbasi {\it et al.}, Phys. Lett. B {\bf 619},
    271 (2005), astro-ph/0501317

\bibitem{hr_prl} R. U. Abbasi {\it et al.}, Phys. Rev. Lett. {\bf 92},
  151101 (2004), astro-ph/0208243

\bibitem{corsika} D. Heck, J. Knapp, J. N. Capdevielle, G. Schatz and
  T. Thouw,
  FZKA 6019 (1998), Forschungszentrum Karlsruhe,\\
  http://www-ik3.fzk.de/\verb!~!heck/corsika/physics\_description/corsika\_phys.html

\bibitem{qgsjet} N. N. Kalmykov, S. S. Ostapchenko and A. I. Pavlov,
  Nucl. Phys. B (Proc. Suppl.) {\bf 52B}, 17 (1997)

\bibitem{hr_hr2} R. U. Abbasi {\it et al.}, Astropart. Phys. {\bf 23}, 157
  (2005), astro-ph/0208301

\bibitem{sibyll} R. S. Fletcher, T. K. Gaisser, P. Lipari and T.
  Stanev, Phys. Rev. D, {\bf 50}, 5710 (1994)
  
\bibitem{hr_atmos1} R. U. Abbasi {\it et al.}, Astropart. Phys. {\bf 25}, 74 (2006)

\bibitem{cowan} G. Cowan, Statistical Data Analysis, Oxford Science
  Publications, 1998

\bibitem{flyseye} D. J. Bird {\it et al.}, Phys. Rev. Lett. {\bf 71}, 3401
  (1993)

\bibitem{ghfunc} T. K. Gaisser and A. M. Hillas, Proceedings of the
  15th International Cosmic Ray Conference, {\bf 8}, 353 (1977)

\bibitem{song} C. Song {\it et al.}, Astropart. Phys. {\bf 14}, 7 (2000),
  astro-ph/9910195

\bibitem{egs} W. R. Nelson, H. Hirayama and D. W. O. Rogers, SLAC
  Report, {\bf 265} (1985)

\bibitem{hadmod} D. Heck, M. Risse and J. Knapp, Nucl. Phys. B (Proc.
  Suppl.) {\bf 122}, 364 (2003)

\bibitem{ostapchenko} S. S. Ostapchenko, presentation at the VIHKOS CORSIKA
school 2005 in Lauterbad, Germany

\bibitem{hrmia1} T. Abu-Zayyad {\it et al.}, Astrophys. J. {\bf 557}, 686
  (2001), astro-ph/0010652

\bibitem{hrmia2} T. Abu-Zayyad {\it et al.}, Phys. Rev. Lett. {\bf 84}, 
  4276 (2000)

\bibitem{hr_stecomp} R. U. Abbasi {\it et al.}, Astrophys. J. {\bf 622},
  (2005) 910-926

\bibitem{fedorova} Y.Fedorova (HiRes Collaboration), Proc. of the 29th 
Int. Cosmic Ray Conf. (2005)

\bibitem{2knee_fe} D. J . Bird {\it et al.}, Ap. J. {\bf 424}, 491 (1994)

\bibitem{2knee_hrmia}  T. Abu-Zayyad {\it et al.}, P.R.L. {\bf 84},  4276 (2000)

\bibitem{2knee_akeno}  N. Nagano {\it et al.}, J. Phys. G {\bf 18}, 423 (1992) 

\bibitem{2knee_yakutsk}  M. I. Pravdin {\it et al.}, Proc. 26th ICRC, Salt Lake City (1999).
  
\bibitem{hr_atmos2} R. U. Abbasi {\it et al.}, Astropart. Phys. {\bf 25}, 93 (2006)  



\end{thebibliography}
\end{document}